\documentstyle[prl,aps,twocolumn,epsf]{revtex}

\newcommand{\ep}{\epsilon}
\begin{document}
\title{Spectral statistics of disordered metals in the presence of several Aharonov-Bohm fluxes}
 \author{G.
Montambaux}
\address{Laboratoire de Physique des Solides,  associ\'e au
CNRS,
 Universit\'{e} Paris--Sud,
 91405 Orsay, France}

\twocolumn[
\maketitle
\widetext
 \begin{center}
\centerline{\today}
 \begin{abstract}
\parbox{14cm}{
The form factor for spectral correlations in a diffusive metal is
 calculated in the presence of several Aharonov-Bohm fluxes.
 When the fluxes $\phi$ are equal, the correlations are universal
functions
of $n g^2   \phi$ where $g$ is the dimensionless conductance and $n$
is the number of applied fluxes. This explains recent flux
dependence of the correlations found numerically at the
 metal-insulator transition.
}
 \end{abstract}
\end{center}
]
\narrowtext

In the recent past, there has been several studies
 on the effect of a Aharonov-Bohm flux on the spectral
properties of disordered metals. These works were
 mainly stimulated by the observation of persistent
 currents in mesoscopic rings\cite{PC}.
In addition, the flux  can be used as a tool to probe the evolution of
the
statistical properties of spectra between different symmetry
 classes\cite{Dupuis91,Pandey83,Akkermans92b,Altland93b}. When no flux is
applied, time
reversal symmetry is preserved, the low energy spectral correlations
are those of an ensemble of
real random matrices\cite{Mehta91,Efetov83,Altshuler86}. This ensemble is
invariant
under orthogonal transformations and is called the   Gaussian Orthogonal Ensemble (GOE).
In a finite flux, time reversal symmetry is broken and the relevant ensemble
of random matrices is the Gaussian Unitary Ensemble (GUE). One or several
magnetic fluxes drive the transition
between the two ensembles. A magnetic flux is also an
example of external parameter which can probe the parametric
correlations, i.e. correlations of energy levels between
different values of a parameter\cite{Simons93}.

Quite recently, a Aharonov -Bohm flux has been applied in numerical
 experiments to study the transition between universality classes at
 the metal-insulator transition\cite{Batsch96} . In this work, several
 fluxes were applied in perpendicular directions and the authors
 noticed that the cross over depends on the number of applied fluxes.

The aim of this short note is to describe analytically the effect of several fluxes. Namely, I
extend the semi-classical calculation of the spectral correlations in
metals, and in particular of the form factor, to the case where  several
fluxes are applied.
Consider  a three-dimensional disordered metal in the diffusive regime, i.e.
when all the dimensions $L_x$, $L_y$ and $L_z$ are larger than the mean free
path $l_e$, so that the electronic motion is diffusive in the three
directions.
Periodic boundary conditions are imposed in the three directions but the
hypertorus can be pierced by  magnetic fluxes $\phi_x, \phi_y, \phi_z$ in
three perpendicular
directions. It is known that the effect of a flux line is simply to change the boundary conditions so that:
\begin{eqnarray}
\psi(x+L_x,y,z)&=&e^{i 2 \pi \varphi_x} \psi(x,y,z) \\
\psi(x,y+L_y,z)&=&e^{i 2 \pi \varphi_y} \psi(x,y,z)   \\
\psi(x,y,z+L_z)&=&e^{i 2 \pi \varphi_z} \psi(x,y,z)
\end{eqnarray}
where $\varphi_i  = \phi_i /\phi_0$, with $i=x,y,z$. $\phi_0$ is the flux quantum.

To describe the spectral correlations, I use the semi-classical result found by Argaman {\it et al.}\cite{Argaman93}  to describe the two-point correlation
 of the density of states $K(\ep) = \langle \rho(E) \rho(E+\ep) \rangle - \langle \rho(E) \rangle^2$.
These authors have related the form factor $\tilde K(t)$, the Fourier transform
of this two-point correlation function ($\hbar=1$ throughout the paper):
\begin{equation} \tilde K (t) = {1 \over 2 \pi} \int K(\ep) \exp(i \ep t) dt
\end{equation}
to the return probability $P(t)$ for a diffusive particle.
In the presence of magnetic fluxes $\varphi_i$, this relation is:
\begin{equation}\tilde K(t,\varphi_i)= t P(t,\varphi_i)/(4 \pi^2)\ \ \ .\end{equation}

When the phase of electrons is coherent, the return probability contains two terms,
the classical return probability $P_{cl}$
 and an interference term, $P_{int}$, which results from phase
 coherence between a diffusive path and the same path obtained
 by time reversal symmetry. The latter term depends on the magnetic
 fluxes. It is the solution of the diffusion equation and is given by:
\begin{equation}P_{int}(t,\varphi_i)= \sum_{\{q_x,q_y,q_z\}} e^{-D (q_x^2+q_y^2+q_z^2) t}
\label{diff}
\end{equation}
where the wave vectors of the diffusion modes are determined by
 the boundary conditions: $q_i= 2  (n_i + 2\varphi_i) \pi /L_i$
 and $n_i =0,\pm 1, \pm2, \cdots$. $D$ is the diffusion
coefficient. For our purpose,
it is more convenient to rewrite this interference term as a sum over winding numbers around each flux tube:

 \begin{eqnarray}
P_{int}(t,\varphi_i)&=&{L_x L_y L_z \over (4 \pi D t)^{3/2}}\sum_{\{m_i\}=-\infty}^{\infty}e^{{-
 \sum_{i}  (m_i^2 L_i^2)  / 4 D t}}  \nonumber     \\
&\times & \cos (4 \pi  \sum_{i} m_i \varphi_i) \end{eqnarray}

$L_i$ the dimension of the system
in the direction $i$.
The sums are over $i=x,y,z$. This expression can be
deduced from eq. \ref{diff} by Poisson summation. Quite simply, this result
expresses that the $m_i^{th}$ harmonic of the flux dependent part of
the return probability  depends on the probability to return to the
origin after $m_i$ windings on the hypertorus.   This is a simple
generalization of the one flux case\cite{Montambaux95}. 
Thus the spectral correlations, in particular the form factor $\tilde K(t)$,
are related to the way the phase is accumulated along
paths of diffusive trajectories after enclosing the different fluxes.

In the limit of
small fluxes $\varphi_i \ll 1$,  the flux periodicity
can be neglected and the above sums can be replaced by gaussian integrals. One
gets:
\begin{displaymath}
P_{int}(t,\varphi_i)=\exp[-16 \pi^2   \sum_i E_{ci} \varphi_i^2 t ]
\end{displaymath}
where $E_{ci}=\hbar D /L_i^2$ is the Thouless energy in the  direction $i$.
The form factor is thus given by:

\begin{displaymath}
\tilde K(t,\varphi_i)={t \over 4 \pi^2} [1 + e^{-16 \pi^2   \sum_i E_{ci} \varphi_i^2 t} ]
\end{displaymath}
The first term of this expression comes from the classical   return probability  which is equal to the interference term when the fluxes are zero.

When all the dimensions and applied fluxes are identical, $\phi_i=\phi$, $E_{ci}=E_c$, a situation considered in ref.\cite{Batsch96},
the form factor depends on the single combination of parameters
$n E_c \varphi^2 t$. This implies that the flux dependence of the two-point
correlation function
is solely governed by $n g \varphi^2$ where $g =
E_c/\Delta$ is the dimensionless conductance,
 $n$ is the number of applied fluxes and $\Delta$ is the mean level spacing.
As expected, the transition between
orthogonal and unitary symmetries is thus faster when three fluxes are applied.

Our calculation has been done in the semiclassical regime where the time is smaller than the Heisenberg time $\tau_H = h / \Delta$.
 The form factor for the orthogonal-unitary transition is
 completely known from Random Matrix theory\cite{Pandey83} or from
the non-linear $\sigma$ model\cite{Altland93b}.
More generally, one expects all correlations functions to be universal functions of the combination $n g \varphi^2$.

Let us now come to the spectral correlation at the metal-insulator transition.
At the transition, the conductance $g$ is scale independent so that the transition between
orthogonal and unitary symmetry classes is expected to be size independent.
This scale invariance has been found numerically recently by Batsch {\it et
al.}\cite{Batsch96}.  In their work, they study the evolution of the distribution $P(s)$ of spacings between nearest energy levels.
They introduce the parameter
$\Gamma_{max}(\varphi)=
(P_m(\varphi)-P_m^o)/(P_m^u-P_m^o)$, where $P_m$ is the maximum of the
 spacing distribution for the orthogonal ($P_m^o$), unitary
    ($P_m^u$) and    intermediate ($P_m(\varphi)$) cases. To describe the
cross-over between orthogonal and unitary symmetries, they present
  different plots of the variation $\Gamma_{max}(\varphi)$,
for the three cases where one, two or three identical  fluxes are applied.
They find that each flux dependence is scale invariant at the transition, in accordance with the scale invariance of the conductance. However, they obtain  three
different variations (fig.3 in ref.\cite{Batsch96}), depending on the number
of applied fluxes, a  result that they qualify as "unexpected behavior".

As for $K(\ep,\varphi)$, one expects the quantity
$\Gamma_{max}(\varphi)$ to be a universal function of $n g \varphi^2$ in the metallic regime. One may now wonder if this scaling also holds at the metal-insulator transition. If it is so,  one
expects that the three curves obtained in fig.3 of
ref.\cite{Batsch96}, for one, two and three fluxes,
merge into a {\it single universal curve} by rescaling $\varphi \rightarrow
\varphi/ \sqrt{n}$.
This is indeed the case as it is shown on fig.1. I have used the
data
 of the fig.3 in  ref.[\cite{Batsch96}]. The flux scales have been renormalized as
$\varphi \rightarrow  \varphi \sqrt{3} $
 for one flux,
 $\varphi
\rightarrow \varphi \sqrt{2}/\sqrt{3}$ for two  fluxes, so that all the curves
now coincide with the three-fluxes curve. This shows that indeed $\Gamma_{max}(\varphi)$ is a unique function of $n \varphi^2$.

It may appear surprising that the scaling found in the diffusive regime still applies at   the Anderson transition. At the transition, the correlation
functions have indeed a different structure. In particular, it has been
found recently
that the form factor is a function of $\varphi^2 t^{1+\eta/d}$ where $\eta$ is
an exponent related to the multifractal structure of the wave function at
the transition\cite{Chalker96}. However, several
fluxes still contribute additively
 to the total accumulated phase so that the
orthogonal-unitary cross-over is still a unique function of $n g^*
\varphi^2$  at the transition, where $g^*$ is scale invariant.

 In conclusion, we have calculated the spectral form factor in the presence
of several Aharonov-Bohm fluxes. When these flux are equal, the
Orthogonal-Unitary transition is driven by the combination of parameters $n
g \varphi^2$. This result still holds at the transition and explains recent
numerical results.

I thank M. Pascaud and P. Walker for useful comments.

 \begin{figure}
\centerline{
\epsfxsize 8cm
\epsffile{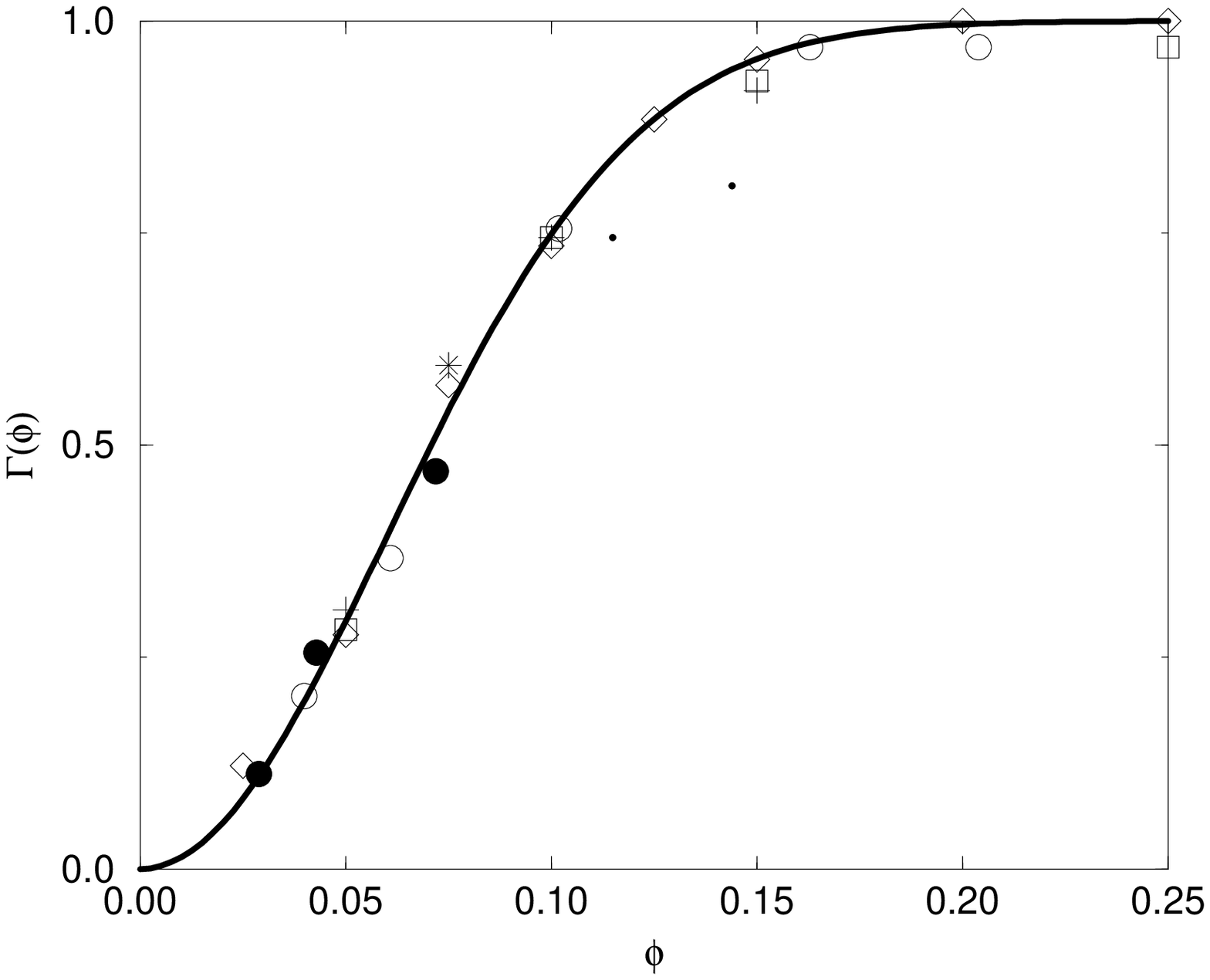}}
\caption{The universal function $\Gamma_{max}(\varphi)$ after the rescaling
proposed in the text. The  two small dots, which correspond to the case of
one flux
are outside the universal curve, because  the transition cannot be
completed before the flux periodicity.
 For the legend of the symbols, see ref.\protect\cite{Batsch96}. The continuous line is a fit to
the data. } \label{fig} \end{figure}

\end{document}